\documentclass[nolineno]{gd-llncs}

\usepackage{graphicx}

\usepackage[hidelinks]{hyperref}
\usepackage{xcolor}

\usepackage{ragged2e}
\usepackage{multirow}
\usepackage{mathtools}
\usepackage{cite} %
\usepackage{csquotes}
\usepackage{placeins}
\usepackage{amsmath}
\usepackage{amssymb}
\usepackage{marvosym}
\usepackage{subcaption}
\graphicspath{{./}} %

\newcommand{\inappendix}[2]{#1}

\newcommand{\crn}{\mathit{cr}}
\newcommand{\ssep}{\mid}

\newcommand{\contr}[1]{\hat{#1}}

\newcommand{\bigO}{\mathcal{O}}
\newcommand{\dual}[1]{{#1}^*}
\newcommand{\edge}[2]{({#1},{#2})}
\newcommand{\ctimes}{\mathbin{\text{\scalebox{.84}{$\square$}}}}

\newcommand{\myparagraph}[1]{\paragraph{#1}}

\begin{document}

\title{Star-Struck by Fixed Embeddings:\\ Modern Crossing Number Heuristics%
\thanks{Supported by the German Research Foundation (DFG) grant CH 897/2-2}}

\titlerunning{Star-Struck by Fixed Embeddings}

\author{Markus~Chimani\orcidID{0000-0002-4681-5550} \and
Max~Ilsen$^{\text{\Letter}}$\orcidID{0000-0002-4532-3829} \and
Tilo~Wiedera\orcidID{0000-0002-5923-4114}}

\institute{Theoretical Computer Science, Osnabrück University, Osnabrück, Germany\\
\email{\{markus.chimani,max.ilsen,tilo.wiedera\}@uos.de}}

\authorrunning{M. Chimani, M. Ilsen, and T. Wiedera}

\maketitle

\begin{abstract}
    We present a thorough experimental evaluation of several crossing
    minimization heuristics that are based on the construction and iterative
    improvement of a planarization, i.e., a planar representation of a graph
    with crossings replaced by dummy vertices.
    The evaluated heuristics include variations and combinations of the
    well-known planarization method, the recently implemented star reinsertion
    method, and a new approach proposed herein: the mixed insertion method.
    Our experiments reveal the importance of several implementation details such
    as the detection of non-simple crossings (i.e., crossings between adjacent
    edges or multiple crossings between the same two edges).
    The most notable finding, however, is that the insertion of stars in a fixed
    embedding setting is not only significantly faster than the insertion of
    edges in a variable embedding setting, but also leads to solutions of higher
    quality.
\end{abstract}

\section{Introduction}
Given a graph $G$, the \emph{crossing number} problem asks for the minimum
number of edge crossings in any drawing of $G$, denoted by $\crn(G)$.
This problem is NP-complete~\cite{garey1983crossing}, even when $G$ is restricted to
cubic graphs~\cite{DBLP:journals/jct/Hlineny06a} or graphs that become
planar after removing a single edge~\cite{DBLP:journals/algorithmica/CabelloM11}.
While the currently known integer linear programming approaches to the
problem~\cite{DBLP:conf/esa/ChimaniW16, DBLP:journals/disopt/BuchheimCEGJKMW08,DBLP:conf/esa/ChimaniMB08}
solve sparse instances within a reasonable time
frame~\cite{DBLP:journals/jea/ChimaniGM09}, dense instances require the use of
heuristics.

One such heuristic is the well-known \emph{planarization
method}~\cite{DBLP:journals/jss/BatiniTT84,DBLP:conf/gd/GutwengerM03}, which
constructs a \emph{planarization}, i.e., a planar representation of $G$
with crossings replaced by dummy vertices of degree $4$.
The heuristic first computes a spanning planar subgraph of $G$ and then
iteratively inserts the remaining edges. %
Several variants of the planarization method have been thoroughly evaluated,
including different edge insertion algorithms and postprocessing
strategies; see~\cite{DBLP:journals/jgaa/ChimaniG12} for the latest study.
In a recent paper~\cite{DBLP:journals/jgaa/ClancyHN19}, Clancy et al.\ present
an alternative heuristic---the \emph{star reinsertion method}---, which differs
in two key aspects from the planarization method: It (i) starts with a full
planarization (instead of a planar subgraph) that is iteratively improved by
reinserting elements, and (ii) the reinserted elements are stars (vertices with
their incident edges) rather than individual edges.
These star insertions are performed using a straight-forward but never tried
algorithm from literature~\cite{DBLP:conf/soda/ChimaniGMW09}.
Clancy et al.\ were faced with the problem that the implementations of the
aforementioned heuristics were written in different languages, leading to
incomparable running times.
In their evaluation, they thus focus on variants of the star reinsertion method;
their comparison with the planarization method only gives averages over (a quite
limited number of) full instance sets and relies on old data from previous
experiments.

Herein, we present a comprehensive experimental evaluation of a wide
array of crossing minimization heuristics based on edge and star insertion
encompassing all known strong candidates.
This includes not only variants of the planarization and star reinsertion
methods but also \emph{combined} approaches.
In addition, we present and evaluate a new heuristic that builds up a
planarization from a planar subgraph using \emph{both} star and edge insertions.
All of these algorithms are implemented as part of the same framework, enabling
us to accurately compare their running times.
Furthermore, we suggest ways of simplifying the implementation of the
heuristics, increasing their speed in practice, and improving their
results---e.g., by properly handling crossings between adjacent edges and
multiple crossings between the same two edges.

\section{Preliminaries}\label{sec:preliminaries}
In the following, we consider a connected undirected graph $G$ (that is usually
simple, i.e., does not contain parallel edges or self-loops) with $n$ vertices
and $m$ edges, denoted by $V(G)$ and $E(G)$ respectively.
Let $\Delta$ be the maximum degree of any vertex in $V(G)$ and $N(v) \coloneqq
\{w \ssep \edge{v}{w} \in E\}$ the neighborhood of a vertex~$v$.
Then, $v$ along with a subset of its incident edges $F \subseteq \{\edge{v}{w}
\in E\}$ is collectively called a \emph{star}, denoted by~$(v,F)$.
Furthermore, a (combinatorial) \emph{embedding} of a planar graph $G$
corresponds to a cyclic ordering of the edges around each vertex in $V(G)$ such
that the resulting drawing can be realized without any edge crossings.
This induces a set of cycles that bound the \emph{faces} of the embedding.
Based on a combinatorial embedding of the \emph{primal graph}~$G$, we can define
the \emph{dual graph}~$\dual{G}$, whose vertices correspond to the faces of~$G$,
and vice versa.
For each primal edge $e \in E(G)$, there exists a dual edge $\dual{e} \in
E(\dual{G})$ between the dual vertices corresponding to the $e$-incident primal
faces.
Note that $\dual{G}$ may be a multi-graph with self-loops even if $G$ is~simple.

For the purpose of this paper, it is of particular concern how to insert an edge
$\edge{v_1}{v_2}$ into a planarization.
First, it is necessary to find a corresponding \emph{insertion path}, i.e.,
a sequence of faces $f_1,\dots,f_k$ such that $v_1$ is incident to $f_1$, $v_2$
incident to $f_k$, and $f_i$ adjacent to $f_{i+1}$ for $i \in \{1,\dots,k-1\}$.
An edge between $v_1$ and $v_2$ can then be inserted into a planarization by
subdividing a common edge for each face pair $(f_i,f_{i+1})$ and routing the
new edge as a sequence of edges from $v_1$ along the subdivision vertices to
$v_2$.
By extension, the \emph{insertion spider} of a star $(v, F)$ is a set of
insertion paths, one for each edge in $F$. These insertion paths necessarily
share a common face into which $v$ can be inserted.

\section{Algorithms}
\label{sec:algorithms}

\subsection{Solving Insertion Problems}
\label{subsec:insertion_problems}
Insertion problems, and their efficient solutions, form the cornerstone of all
known strong crossing minimization heuristics.

\begin{definition}[EIF, SIF]
    Given a planar graph~$G$, an embedding~$\Pi$ of $G$, and an edge (or star)
    not yet in~$G$, insert this edge (star) into $\Pi$ such that the number of
    crossings in $\Pi$ is minimized.
    We refer to these problems as the \emph{edge (star) insertion
    problem with fixed embedding~EIF~(SIF,~resp.)}.
\end{definition}

Given a primal vertex $v$, let $\contr{v}$ be the vertex that is created by
contracting the dual vertices that correspond to $v$-incident faces.
Then, the EIF for any given edge~$\edge{v_1}{v_2}$ can be solved optimally in
$\bigO(n)$ time by computing the shortest path from $\contr{v_1}$ to
$\contr{v_2}$ in the dual graph~$\dual{G}$ via breadth-first
search~(BFS)~\cite{DBLP:journals/jss/BatiniTT84}.
By extension, the SIF for a star $(v,F)$ can be solved in $\bigO(|F| \cdot n)$
time as follows~\cite{DBLP:conf/soda/ChimaniGMW09}:
For each edge $(v,w) \in F$, solve the single-source shortest path problem in
$\dual{G}$ with $\contr{w}$ as the source (via BFS).
For each face $f$, the sum over all of the resulting distance values at this $f$
then represents the number of crossings that would be created if $v$ was to be
inserted into $f$.
Hence, the face with the minimum distance sum is the optimal face to
insert $v$ into, and the computed shortest paths to this face collectively
form the insertion spider.
To avoid crossings between these shortest paths (due to them not being
necessarily unique), we can construct the insertion spider using a final
BFS starting at the optimal face.

\begin{definition}[EIV, MEIV, SIV]
    Given a planar graph~$G$ and an edge (a set of $k$~edges, or a star) not yet
    in $G$, find an embedding~$\Pi$ among all possible embeddings of~$G$ such
    that optimally inserting the edge (set of $k$~edges, star) into this~$\Pi$
    results in the minimum number of crossings.
    We refer to these problems as the \emph{edge (multiple edge, star) insertion
    problem with variable embedding~EIV~(MEIV,~SIV,~resp.)}.
\end{definition}

The EIV can be solved in $\bigO(n)$~time using an algorithm by Gutwenger et
al.~\cite{DBLP:journals/algorithmica/GutwengerMW05}, which finds a suitable
embedding (with the help of SPR-trees) and then executes the EIF-algorithm
described above.
Now consider the MEIV: Solving it for general $k$ is
NP-hard~\cite{DBLP:phd/dnb/Ziegler01}, however there exists an $\bigO(kn +
k^2)$-time approximation algorithm with an additive guarantee of $\Delta k \log
k + \binom{k}{2}$~\cite{DBLP:journals/jco/ChimaniH17} that performs well in
practice~\cite{DBLP:journals/jgaa/ChimaniG12}.
Put briefly, the EIV-algorithm is run for each of the $k$~edges independently,
and a single final embedding is identified by combining the individual
(potentially conflicting) solutions via voting.
Then, the EIF-algorithm can be executed once for each edge.
Note that the SIV can be solved optimally in polynomial time by using
dynamic programming techniques~\cite{DBLP:conf/soda/ChimaniGMW09}.
However, for graphs that are not series-parallel, the resulting running times
are exorbitant and there is no known implementation of this algorithm.
In fact, our results herein suggest that in the context of crossing minimization
heuristics, the solution power of the SIV-algorithm is fortunately not
necessary in practice.

Each problem discussed above has a \emph{weighted} version which can be solved
in the same manner if each $c_e$-weighted edge $e$ is replaced by $c_e$ parallel
$1$-weighted edges beforehand.
In practice it is worthwhile to compute the shortest paths during the
EIF/SIV-algorithm on the weighted instance directly. However, this does not
allow for the same theoretical upper bounds of the running times
since the weights may be arbitrarily large.

\subsection{Crossing Minimization Heuristics}
\label{subsec:heuristics}
We start with reviewing several crossing minimization heuristics that
iteratively build up a planarization, starting with a planar subgraph:

\myparagraph{The planarization method (plm)} is the longest studied and best-known
approach considered, achieving strong results in previous
evaluations~\cite{DBLP:journals/jss/BatiniTT84,DBLP:journals/jgaa/ChimaniG12,DBLP:conf/gd/GutwengerM03}.
First, we compute a spanning planar subgraph $G' = (V,E') \subseteq G$, usually
by employing a maximum planar subgraph heuristic and extending the result such
that it becomes (inclusion-wise) maximal.
Then, the remaining edges $F \coloneqq E \setminus E'$ are either inserted one
after another---by solving the respective EIF (\emph{fix}) or EIV
(\emph{var})---or simultaneously using the MEIV-approximation algorithm
(\emph{multi}).
Gutwenger and Mutzel~\cite{DBLP:conf/gd/GutwengerM03} describe a postprocessing
strategy for \emph{plm} based on edge insertion: Each edge is deleted from the
planarization and reinserted one after another~(\emph{all}).
To incrementally improve the planarization, \emph{all} can also be executed once
after each individual edge
insertion~(\emph{inc})~\cite{DBLP:journals/jgaa/ChimaniG12}.
In the following, we represent the use of these postprocessing strategies by
appending the respective shorthand to the algorithm's abbreviation, e.g.\
\emph{fix-all}.
When neither \emph{all} nor \emph{inc} is employed, we use the specifier
\emph{none} instead.

\myparagraph{The chordless cycle method (ccm)} realizes the idea of extending a
\emph{vertex-induced} planar subgraph to a full planarization via star
insertion~\cite{DBLP:conf/soda/ChimaniGMW09}. It corresponds to the
best-performing scheme for the star insertion algorithm as examined by Clancy et
al.~\cite{DBLP:journals/jgaa/ClancyHN19}:
Search for a chordless cycle in $G$, e.g., via breadth-first search. Let $G'$
denote the subgraph of $G$ that is already embedded and initialize it with this
chordless cycle.
Iteratively (until the whole graph is embedded) select a vertex $v \not\in
V(G')$ such that there exists at least one edge $\edge{v}{w}$ that connects $v$
with the already embedded subgraph $G'$; insert $v$ into $G'$ by solving the SIF
for the star~$(v,~\{\edge{v}{w} \in E \ssep w \in V(G')\})$.

\myparagraph{The mixed insertion method (mim)} is a novel approach that we propose
as an alternative to the planarization schemes above. It proceeds in a fashion
that is similar to \emph{plm} but relies on star insertion instead of edge
insertion in as many cases as possible.
Accordingly, let $G'$ denote the subgraph of $G$ that is already embedded and
initialize it with a spanning planar subgraph $(V,E') \subseteq G$.
Then, (attempt to) insert the remaining edges $F \coloneqq E \setminus E'$ by
reinserting at least one endpoint of each edge $e \in F$ via star insertion.
Since removing and then reinserting a \emph{cut vertex} of the planar subgraph
$G'$ would temporarily disconnect it, the cut vertices of the planar subgraph
are computed (cf.\ \cite{DBLP:journals/cacm/HopcroftT73}) and each edge $e \in
F$ is processed as follows:
If both endpoints of $e$ are cut vertices of $G'$, insert the edge via edge
insertion (we choose to do so in a variable embedding setting as such edge
insertions happen rarely).
If only one endpoint of the edge is a cut vertex, reinsert the other one.
If neither endpoint of the edge is a cut vertex, the endpoint to be reinserted
can be chosen freely---globally, this corresponds to finding a vertex
cover on the graph induced by $F$ that has to include all vertices
neighboring a cut vertex in $G'$. Finding an optimal vertex cover is NP-hard
\cite{DBLP:conf/coco/Karp72}; therefore we compare several heuristics:
For each edge~$e$, choose one of the endpoints
randomly~(\emph{random}), choose the one with the higher or lower degree
in~$G$~(\emph{high}$_G$, \emph{low}$_G$), choose the one with the higher or
lower degree in the graph induced by all edges in $F$ not incident to a cut
vertex in~$G'$~(\emph{high}$_{F}$, \emph{low}$_{F}$), or choose both
endpoints~(\emph{both}).
Each of the chosen vertices is then deleted from the planar subgraph and
reinserted together with all of its edges in the original graph by solving the
corresponding~SIF.

\myparagraph{}Herein, we evaluate the aforementioned heuristics not only on their own but
also in combination with the \emph{star reinsertion method} (\emph{srm}) by
Clancy et al.~\cite{DBLP:journals/jgaa/ClancyHN19}, a postprocessing strategy
based on star insertion.
It starts with an already existing planarization, which may be constructed using
any of the methods outlined above (or even more trivial ones, such as extracting
a planarization from a circular layout of the vertices, which, however, is known
to perform worse~\cite{DBLP:journals/jgaa/ClancyHN19}).
To represent that the result of an algorithm is improved via \emph{srm}, we
append \enquote{\emph{srm}} to its abbreviation, e.g. \emph{fix-none-srm}.
The given planarization is thereby processed as follows:
Iteratively choose a vertex $v$, delete $v$ from $G$, and reinsert it
again by solving the SIF for the star $(v,v \times N(v))$.
Continue the loop until there is no more vertex whose reinsertion improves
the solution (in which case the latter is said to be \emph{locally
optimal}).
Clancy et al.\ propose different methods for choosing $v$; here, we
consider the scheme they report to be the best compromise between
solution quality and running time:
In each iteration, try to reinsert every vertex once and continue with
the next iteration as soon as a vertex is found whose reinsertion
improves the number of crossings in the~planarization.

The original algorithm only updates a planarization once an actual improvement
is found and resets it to its original state otherwise. We propose to never
reset it.
This approach is permissible as the SIF is solved optimally and the number of
crossings hence never increases after the reinsertion of a star.
Not resetting the planarization has the potential to save time in practice as it
allows for a simpler implementation without any need to copy the dual graph.

\section{A Note on Non-simple Crossings}

\begin{figure}[p]
\begin{minipage}{\textwidth}
\centering
    \begin{subfigure}[b]{0.45\textwidth}
        \centering
        \includegraphics{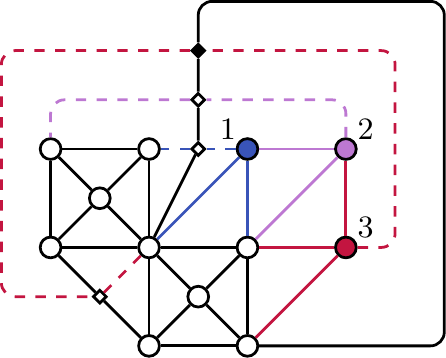}
        \caption{Creation of an $\alpha$-crossing}
    \end{subfigure}\hfill
    \begin{subfigure}[b]{0.45\textwidth}
        \centering
        \includegraphics{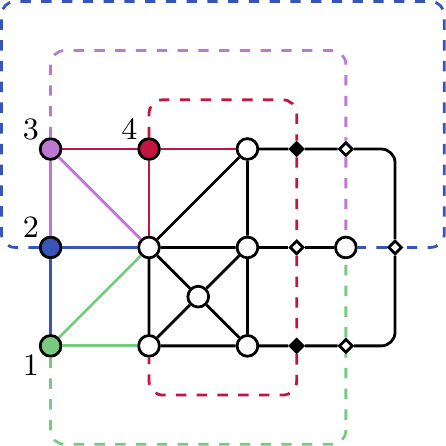}
        \caption{Creation of a $\beta$-crossing}
    \end{subfigure}
\end{minipage}
\caption{A non-simple crossing on the red dashed edge as the result of
    incrementally solving the same kind of insertion problem.
    When starting with the black planar subgraph, this may happen by solving the
    SIV using the described algorithm for the colored vertices in the order of
    their label numbers.
    Alternatively, if all solid edges constitute the initial planar subgraph,
    solving the EIV for the dashed edges in the order of their label numbers
    can have the same result.
    The examples apply both in the fixed and the variable embedding setting.
    Dummy vertices for (non-simple) crossings are represented by small (black)
    diamonds.
}
\label{fig:crossing_existence}
\end{figure}

\begin{figure}[p]
\begin{minipage}{\textwidth}
\centering
    \begin{subfigure}[b]{0.45\textwidth}
        \centering
        \includegraphics{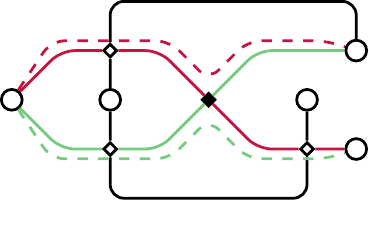}
        \caption{Removal of an $\alpha$-crossing}
    \end{subfigure}\hfill
    \begin{subfigure}[b]{0.45\textwidth}
        \centering
        \includegraphics{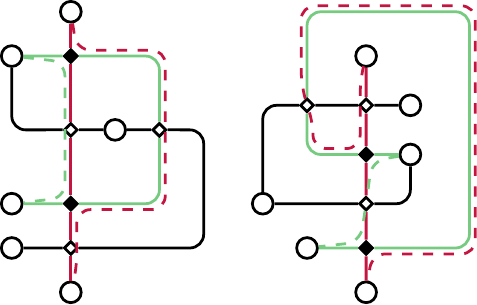}
        \caption{Removal of a $\beta$-crossing}
    \end{subfigure}
\end{minipage}
\caption{Non-simple crossings between the red and green edges. After their
removal (new edge paths drawn as dashed), the red edge is involved in a
new~non-simple crossing of the same type and the green edge in a new non-simple
crossing of the opposite type.
Thus, the removal procedure may have to be iterated.}
\label{fig:non_simple_crossings}
\end{figure}

It is well-known that any crossing-optimal drawing can be assumed to be
\emph{simple}: No edge self-intersects and each pair of edges intersects at most
once (either in a crossing or an endpoint).
In particular, a simple drawing may not contain crossings between adjacent edges
(\emph{$\alpha$-crossings}) or multiple crossings between the same two edges
(\emph{$\beta$-crossings}).
We may hence call any such undesired crossings \emph{non-simple}.
Surprisingly, earlier implementations of the planarization method did not
consider the emergence and removal of any non-simple
crossings~\cite{DBLP:journals/jgaa/ChimaniG12} while the implementation of the
star reinsertion method by Clancy et al.\ only considers $\beta$- but not
$\alpha$-crossings~\cite{DBLP:journals/jgaa/ClancyHN19}.
However, we show in Figure~\ref{fig:crossing_existence} that incrementally
solving the same kind of insertion problem may result in a planarization
with $\alpha$- or $\beta$-crossings, even when starting with a planar subgraph.
Non-simple crossings can be removed by reassigning edges in the planarization to
different edges in the original graph and then deleting the respective dummy
vertices (see Figure~\ref{fig:non_simple_crossings}).
Doing so leads to better results overall,
\inappendix{as shown in Appendix~\ref{sec:eval_nonsimple_crossings}}%
{see~\cite[Appendix~C]{chimani2021starstruck}}.

\section{Experiments}
\label{sec:experiments}

\myparagraph{Setup:}
All algorithms are implemented in \texttt{C++} as part of the Open Graph Drawing
Framework (OGDF, \url{www.ogdf.net}, based on the release \enquote{2020.02
Catal\-pa})~\cite{DBLP:reference/crc/ChimaniGJKKM13}, and compiled with GCC 8.3.0.
Each computation is performed on a single physical processor of a
Xeon~Gold~6134~CPU~(3.2~GHz), with a memory limit of 4~GB but no time limit.
All instances and results are available for download at
\url{http://tcs.uos.de/research/cr}. %

\myparagraph{Instances:}
Table~\ref{tab:instances} lists the instance sets used for our
evaluation\,(see
\inappendix{Appendix~\ref{sec:instance_statistics}}%
{\cite[Appx.\,A]{chimani2021starstruck}}
for further statistical analysis).
To enable a proper comparison of the tested algorithms (and potentially
in the future, their competitors), we consider multiple well-known benchmark
sets as well as constructed, random, and real-world instances with
varying characteristics.
These are preprocessed by computing the \emph{non-planar
core}~(NPC)~\cite{DBLP:journals/dm/ChimaniG09} for each non-planar
biconnected component.
We consider only those instances that have at least $25$ vertices after the NPC
reduction unless the instance is part of the Complete, Complete-Bip., or
KnownCR instance sets.
Moreover, we precompute a planar subgraph and chordless cycle for each
instance such that different runs of \emph{plm}, \emph{mim} and \emph{ccm} can
be started with the same initialization.
The planar subgraph is computed by using Chalermsook and Schmid's diamond
algorithm~\cite{DBLP:conf/walcom/Chalermsook017} and extending the result to
a maximal planar subgraph.
On average, this computation took only 0.77\% of the time needed to execute the
fastest evaluated heuristic \emph{fix-none}---a comparatively negligible amount
of time that is not further taken into consideration during the evaluation.

\begin{table}[t]
    \centering
    \caption{Considered instance sets. \enquote{\#} denotes the number of graphs
        and $|V(G)|$ the (range of the) numbers of nodes---both values refer to
        the instance sets \emph{after} preprocessing.
        Further, let $\delta$ denote the node degree, $\ctimes$ the Cartesian
        product of two graphs, $C_i$~the cycle with $i$ edges, $P_j$ the path
        with $j$ edges, and $G_k$ the $21$ non-isomorphic connected graphs on
        $5$ vertices indexed by $k$.
}
    \label{tab:instances}
    \begin{tabular}{lrr@{\ \;}p{7.39cm}}
        Name & \# & $|V(G)|$ & Description \\
        \hline\hline
        \textbf{Rome} & 3668 & 25--58 & Well-known benchmark set~\cite{DBLP:journals/comgeo/WelzlBGLTTV97}, sparse\\
        \hline
        \textbf{North} & 106 &
        25--64 &
        Well-known benchmark set collected by S.\ North~\cite{DBLP:journals/ijcga/BattistaGLPTTVV00}\\
        \hline
        \textbf{Webcompute} & 75 & 25--112 &
        Instances sent to our online tool~\cite{DBLP:conf/esa/ChimaniW16} for the exact computation of crossing
        numbers, \url{crossings.uos.de}\\
        \hline
        \textbf{Expanders} & 240 & 30--100 & 20 random regular graphs~\cite{steger_wormald_1999}
        (\emph{expander graphs} with high probability)
        for each parameterization $(|V(G)|,\delta) \in \{30,50,100\} \times
        \{4,6,10,20\}$\\
        \hline
        \textbf{Circuit-Based} & 45 & 26--3045 &
        \multirow{4}{7.39cm}{\justifying Hypergraphs from real world electrical
        networks, transformed into traditional graphs by replacing each
        hyperedge $h$ by a new hypervertex connected to all vertices contained in $h$}\\
        \emph{ISCAS-85}~\cite{brglez1985neutral} & 9 & 180--3045 & \\
        \emph{ISCAS-89}~\cite{brglez1989notes} & 24 & 60--584 & \\
        \emph{ITC-99}~\cite{DBLP:journals/dt/CornoRS00} & 12 & 26--980 & \\
        \hline
        \textbf{KnownCR} & 1946 & 9--250 & Benchmark set with $\crn$ known through proofs~\cite{gutwenger10}:\\
        $C \ctimes C$ & 251 & 9--250 & $\to$ $C_i \ctimes C_j$ with  $3 \leq i \leq 7$, $j \geq i$ such that $i\cdot j \leq 250$\\
        $G \ctimes P$ & 893 & 15--245 & $\to$ Subset of $G_i \ctimes P_j$ with $1 \leq i \leq 21$, $3 \leq j \leq 49$\\
        $G \ctimes C$ & 624 & 15--250 & $\to$ Subset of $G_i \ctimes C_j$ with $1 \leq i \leq 21$, $3 \leq j \leq 50$\\
        $P(\_,\_)$ & 178 & 10--250 & $\to$ Generalized Petersen graphs $P(2k + 1, 2)$ with $2 \leq k \leq 62$ and $P(m, 3)$ with $9 \leq m \leq 125$\\
        \hline
        \textbf{Complete} & 46 & 5--50 & Complete graphs $K_n$ for $5 \leq n \leq 50$\\
        \hline
        \textbf{Complete-Bip.} & 666 & 10--80 & Complete bipartite graphs\,$K_{n_1,n_2}$\,for\,$5 \leq n_1,n_2 \leq 40$\\
    \end{tabular}
\end{table}

The precomputed chordless cycle almost always consists of 3--6 vertices,
containing 7--11 vertices for only 15~instances overall.
How many edges are deleted to create the planar subgraph, on the other hand,
varies greatly depending on the size and density of the graph.
Of particular interest is the number of deleted edges that are incident to one
or two cut vertices of the planar subgraph:
During \emph{mim}, the former ones have a fixed endpoint that
must be reinserted via star insertion (disallowing a choice of the reinserted
endpoint) while the latter ones must be inserted via edge insertion.
Clearly, more dense instances such as the complete (bipartite) ones and the
expanders require more edges to be deleted to form a planar subgraph.
At the same time, due to their high connectivity, these instances also have less
deleted edges that are connected to cut vertices in the planar subgraph.
In particular, the complete (bipartite) instances do not have a single such edge.
However, even on the sparser instances, \emph{mim} inserts almost all edges via
star insertion and one can usually choose the endpoint to be reinserted (see the
\emph{mim}-variants described in Subsection~\ref{subsec:heuristics}).

\subsection{Fast Heuristics: Mixed Insertion Method, Chordless Cycle Method and
Fixed Embedding Edge Insertion}

\begin{figure}[tbp]
\centering
\begin{subfigure}{\textwidth}
    \centering
    \includegraphics{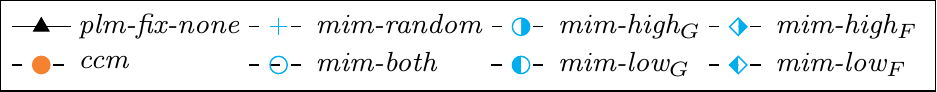}
\end{subfigure}\\[0.1cm]
\begin{subfigure}{.5\textwidth}
    \centering
    \includegraphics{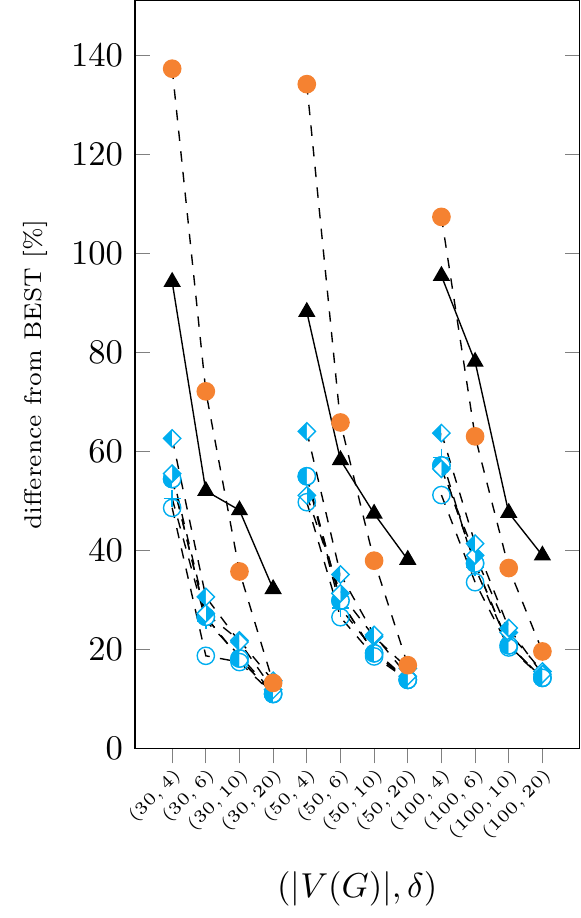}
\end{subfigure}\hfill%
\begin{subfigure}{.5\textwidth}
    \centering
    \includegraphics{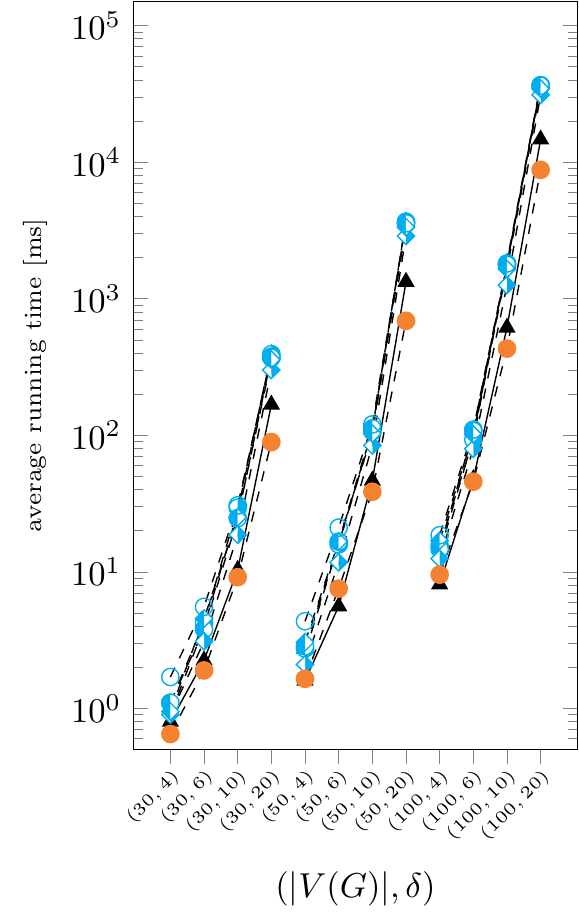}
\end{subfigure}
\caption{Comparison of the \emph{mim}-variants, \emph{ccm} and \emph{fix-none}
on the expanders.}
\label{fig:mim_comparison}
\end{figure}

The \emph{mim}-variants, \emph{ccm}, and \emph{fix-none} (all without
\emph{srm}-postprocessing) are very fast but yield a comparably high number
of crossings.
Figure~\ref{fig:mim_comparison} displays some representative results on the
expanders, contrasting them with the \emph{BEST} solution found by 50
random permutations of any heuristic tested herein (cf.\
Subsection~\ref{subsec:permutations}).
Among the \emph{mim}-variants, there are only little differences in computation
speed and resulting number of crossings.
However, reinserting \emph{both} endpoints whenever a choice between two
endpoints can be made clearly provides the best results across all instances
while only taking an insignificant amount of additional time.
The variant leads to the highest amount of reinserted stars and hence also to
more chances for an improvement of the number of crossings.
In contrast, \emph{high}$_{F}$ needs the lowest amount of star insertions and is
thus the fastest variant (but provides results of mixed quality).

Compared with \emph{fix-none} and \emph{ccm}, \emph{mim} (from now on always
referring to the \emph{both}-variant) provides better results on almost all
instances.
The fastest of the algorithms, on the other hand, is \emph{fix-none}.
The last of the three, \emph{ccm}, should only be considered when examining
particularly dense instances:
On sparse instance sets such as Rome or KnownCR, %
it is slower and yields far worse results than \emph{fix-none} (which in turn
yields worse results than \emph{mim}), but the solution and speed disparity between
the algorithms becomes smaller on instances with a higher density---see, e.g.,
Figure~\ref{fig:mim_comparison}.
On complete (bipartite) instances, \emph{ccm} even surpasses \emph{mim} both in terms
of solution quality and speed.

\subsection{Planarization Method}
\label{subsec:planarization_method}
The different edge insertion algorithms and postprocessing strategies for the
planarization method allow to greatly improve the final planarizations at the
cost of additional running time.
A detailed experimental comparison of these \emph{plm}-variants was already
carried out in 2012~\cite{DBLP:journals/jgaa/ChimaniG12}.
We are able to replicate the results of that study and corroborate its
claims with findings on additional instances:

In terms of solution quality, \emph{none} provides much worse results than
\emph{all} and \emph{inc} across all instance sets.
However, postprocessing and \emph{inc} in particular has the drawback of very
high running times and a large amount of required memory.
Among the edge insertion algorithms, \emph{var} performs better (but is also
slower) than \emph{multi}, which in turn performs better than \emph{fix}.
Overall, \emph{fix-all} is the fastest \emph{plm}-variant that still benefits
from the quality improvements of postprocessing.
The best compromise between solution quality and speed is provided by the
\emph{multi}-variants while the best results are achieved by
\emph{var-inc} (cf.\
\inappendix{Appendix~\ref{sec:plm_appendix}}%
{\cite[Appx.~B]{chimani2021starstruck}}).

\subsection{Improvements via the Star Reinsertion Method}

\begin{figure}[p]
\centering
\begin{subfigure}{\textwidth}
    \centering
    \includegraphics{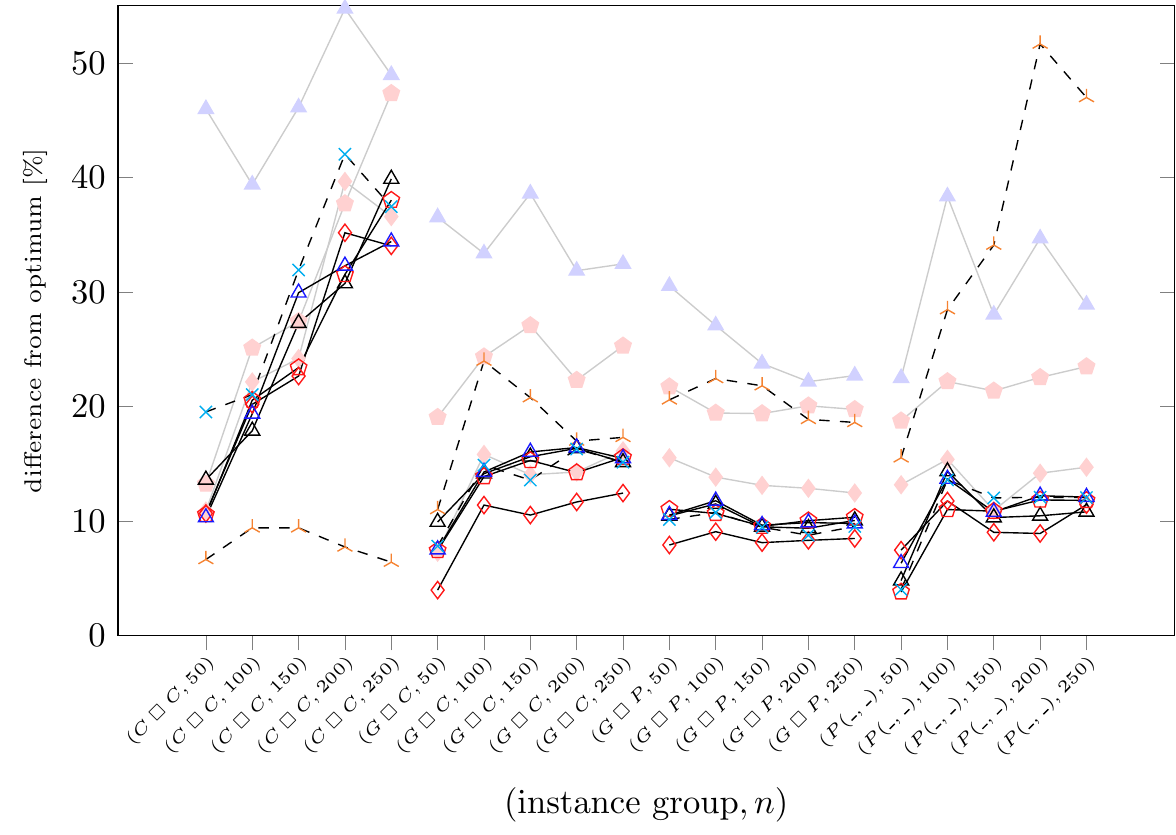}
\end{subfigure}
\\[0.2cm]
\begin{subfigure}{\textwidth}
    \centering
    \includegraphics{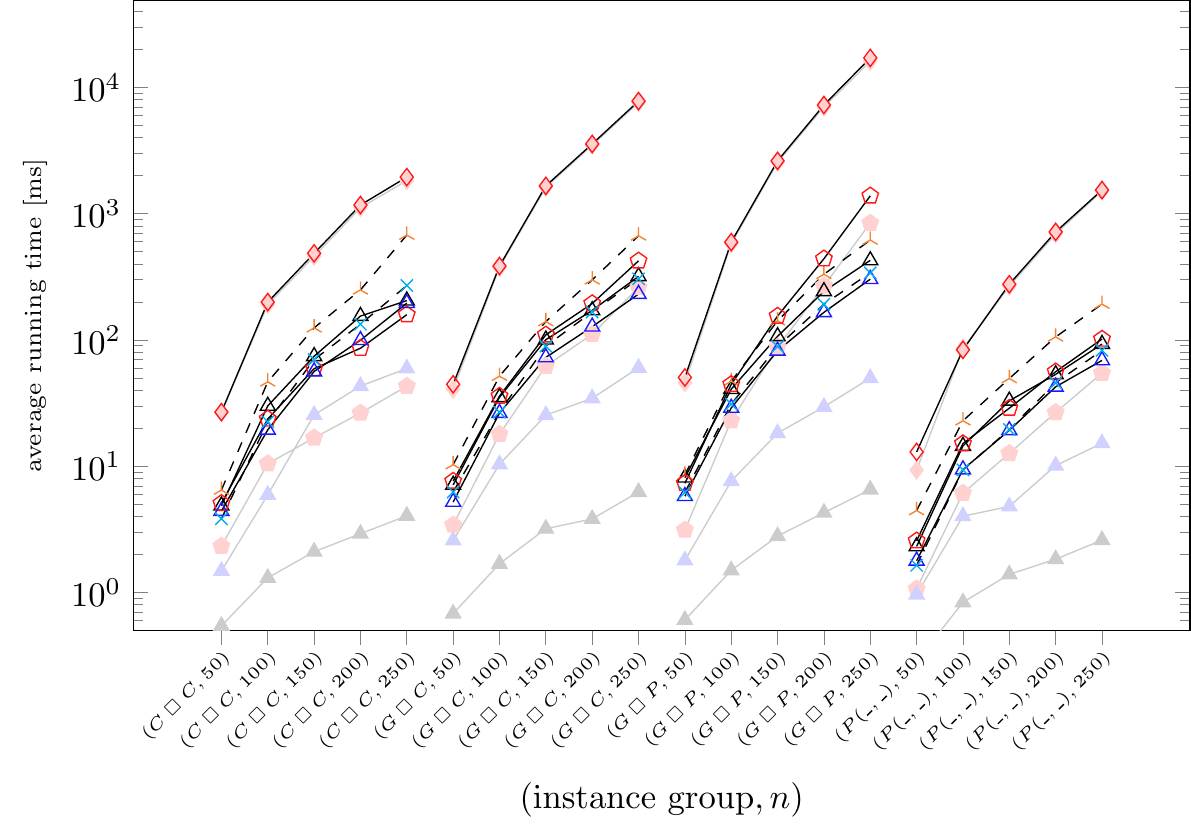}
\end{subfigure}
\caption{Comparison of the \emph{srm}-variants on the KnownCR instances. The
legend of Figure~\ref{fig:srm_comparison} applies. Instance sizes
are rounded up to the nearest multiple of fifty. Note that the results of
\emph{ccm-srm} heavily depend on the structure of the instance; they also vary a
lot across other instance sets\,(with middling results on average).}
\label{fig:srm_comparison_knowncr}
\end{figure}

\begin{figure}[tb]
\centering
\begin{subfigure}{\textwidth}
    \centering
    \includegraphics[width=\textwidth]{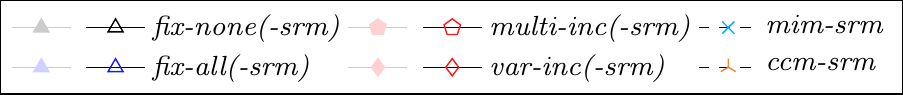}
\end{subfigure}\\[0.2cm]
\begin{subfigure}{.5\textwidth}
    \centering
    \includegraphics[width=\textwidth]{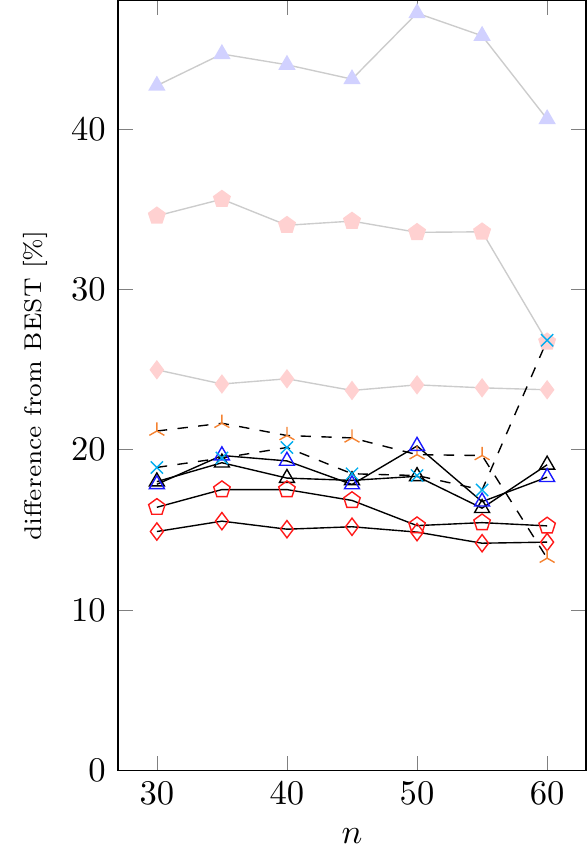}
\end{subfigure}\hfill%
\begin{subfigure}{.5\textwidth}
    \centering
    \includegraphics[width=\textwidth]{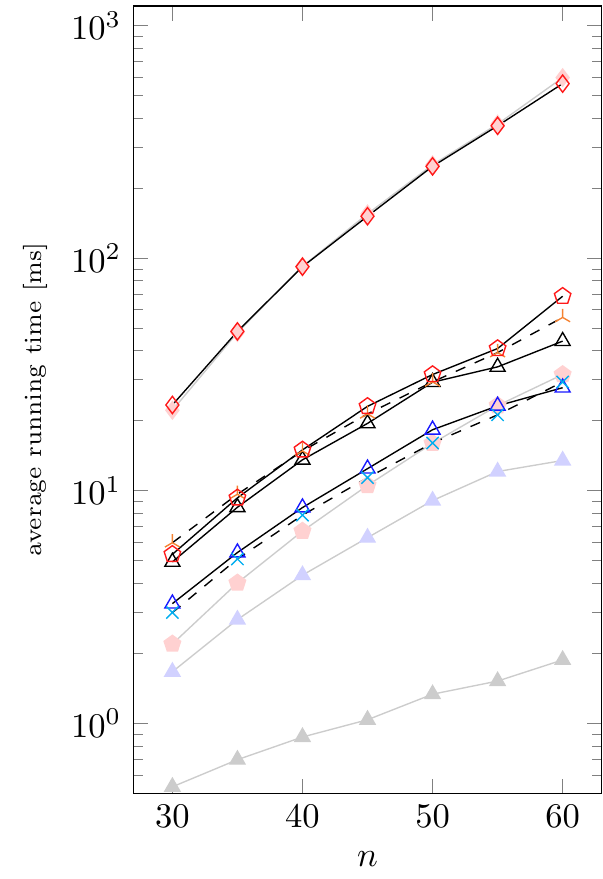}
\end{subfigure}
\caption{Comparison of the \emph{srm}-variants on the Rome instances.
The grayed out plots represent the heuristic variants without
\emph{srm}-postprocessing. Instance sizes are rounded up to the nearest multiple
of five.}
\label{fig:srm_comparison}
\end{figure}

We tested \emph{srm} as a postprocessing method for the eight most promising and
interesting algorithms that construct an initial planarization:
The three fast algorithms \emph{mim}, \emph{ccm}, and \emph{fix-none}, as
well as the more involved \emph{fix-all}, \emph{multi-all}, \emph{multi-inc},
\emph{var-all}, and \emph{var-inc}.
In the case of the latter five, a form of postprocessing is already used, and
the additional application of \emph{srm} only leads to a small increase in
running time, comparatively speaking.
In the case of the former three, the additional postprocessing via \emph{srm}
significantly increases the running times (\emph{fix-none-srm} becomes even
slower than \emph{fix-all-srm}), but the algorithms are still
surprisingly fast:
On sparse instances, the running times are comparable to \emph{multi-inc}
(without \emph{srm}); on dense instances, the algorithms are even faster than
\emph{fix-all}.
This is especially interesting as all \emph{srm}-enhanced algorithms typically
outperform even the best previously known heuristic variant \emph{var-inc} (see
Figures~\ref{fig:srm_comparison_knowncr} and \ref{fig:srm_comparison}).
In spite of its simplicity, star insertion in a fixed embedding setting is able
to greatly improve intermediate planarizations by inserting multiple edges at
once.
It provides better results and is faster than edge insertion in a variable
embedding setting even if the latter uses incremental postprocessing.

When observing the solution quality of the \emph{srm}-algorithms, the same
hierarchy as for the algorithms without \emph{srm} emerges:
\emph{fix-none-srm} performs worse than the other \emph{plm}-based
\emph{srm}-variants, with \emph{var-inc-srm} providing the best results overall.
However, \emph{var-inc-srm} is rarely worth the additional running time since
the three significantly faster \emph{mim-srm}, \emph{ccm-srm} and
\emph{fix-none-srm} perform similarly well or even surpass it on many instances such
as several circuit-based ones and the expanders.
In comparison to \emph{mim-srm} for example, \emph{var-inc-srm}'s
solution quality difference to BEST is only 1.7\% smaller but its median
running time is eight times higher (when averaged over all instances).
The running times of the faster algorithms seem to coincide with the quality
of the planarization delivered by the base algorithm:
While \emph{fix-none-srm} is generally faster than \emph{ccm-srm} on sparse
instances,  %
the opposite is true on denser ones. %
On complete (bipartite) instances, \emph{ccm-srm} becomes even faster than
\emph{mim-srm}. %
However, \emph{mim-srm} is the otherwise fastest among these algorithms, and
thus we recommend to use it.

\subsection{Improvements via Permutations}
\label{subsec:permutations}

We will consider one last question: Whether multiple runs of the same algorithm
with different random permutations of the inserted elements can
significantly improve the results.
For \emph{plm}, we permute the order in which the deleted edges are inserted,
and for \emph{mim}, \emph{ccm} and \emph{srm}, we permute the order of
(re)inserted stars.
Our experiments compare the effect of 50 random permutations with
respect to the Rome, North, Webcompute and KnownCR instance sets.
For the larger instances and more time-consuming algorithms,
this number of permutations is the limit of what we are able to compute.
We focus on the \emph{(relative) improvement} for each instance, i.e., the
lowest number of crossings divided by the average number of crossings across 50
permutations (%
\inappendix{cf.\ Appendix~\ref{sec:relative_improvement}}%
{cf.\ \cite[Appendix~D]{chimani2021starstruck}}).

\begin{figure}[p]
\centering
\captionsetup[subfigure]{justification=centering}
\begin{subfigure}{.5\textwidth}
    \centering
    \includegraphics{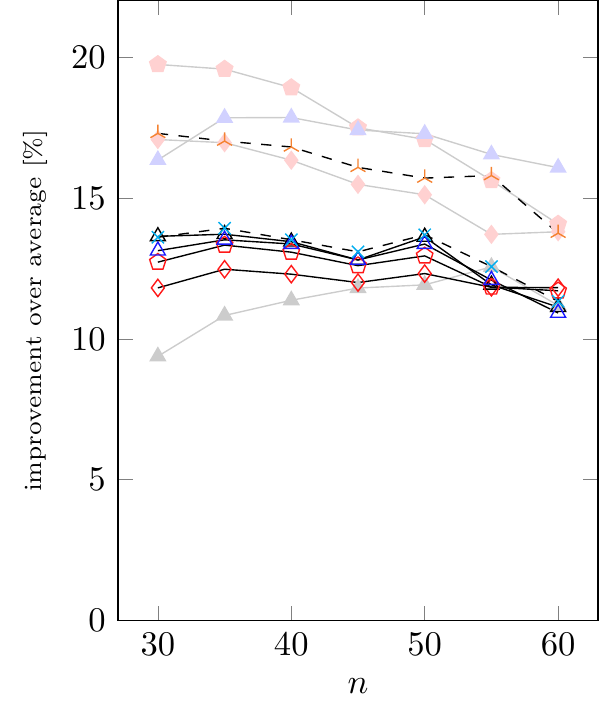}
    \caption{Rome}
\end{subfigure}\hfill%
\begin{subfigure}{.5\textwidth}
    \centering
    \includegraphics{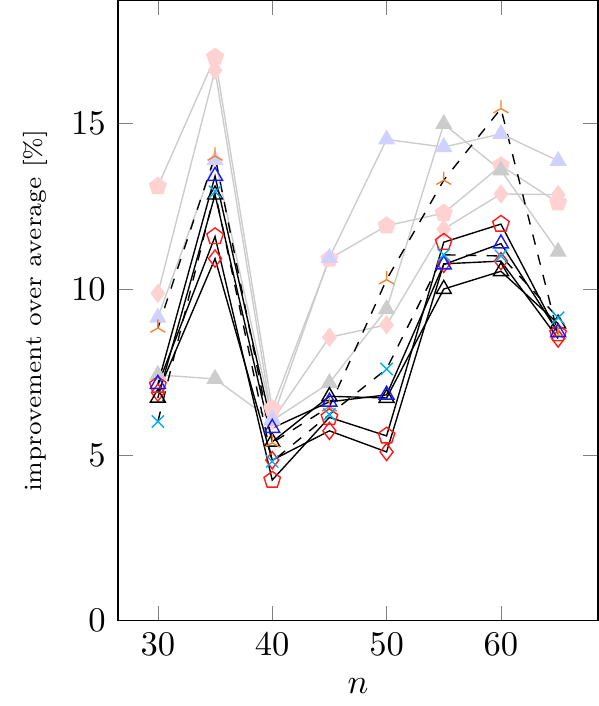}
    \caption{North}
\end{subfigure}%
\caption{Comparison of relative improvements for 50 permutations over their
    average on the Rome and North instances.
    The legend of Figure~\ref{fig:srm_comparison} applies.}
\label{fig:relative_comparison}
\end{figure}

\begin{figure}[p]
\centering
\captionsetup[subfigure]{justification=centering}
\begin{subfigure}{\textwidth}
    \centering
    \includegraphics[width=.98\textwidth]{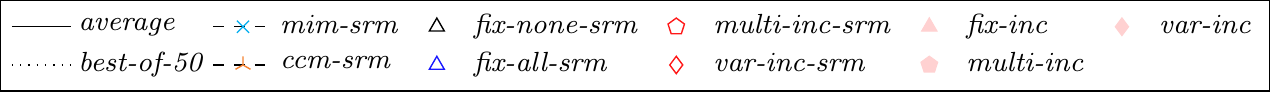}
\end{subfigure}
\\[0.2cm]
\begin{subfigure}{.5\textwidth}
    \centering
    \includegraphics{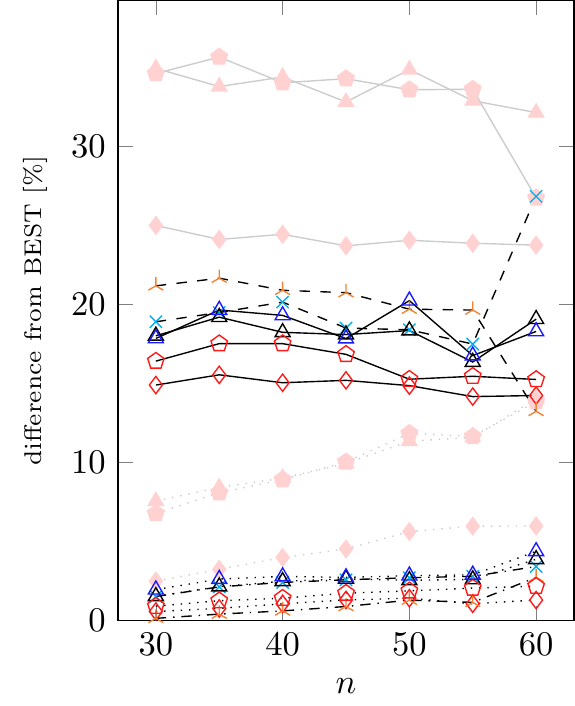}
    \caption{Rome}
\end{subfigure}\hfill%
\begin{subfigure}{.5\textwidth}
    \centering
    \includegraphics{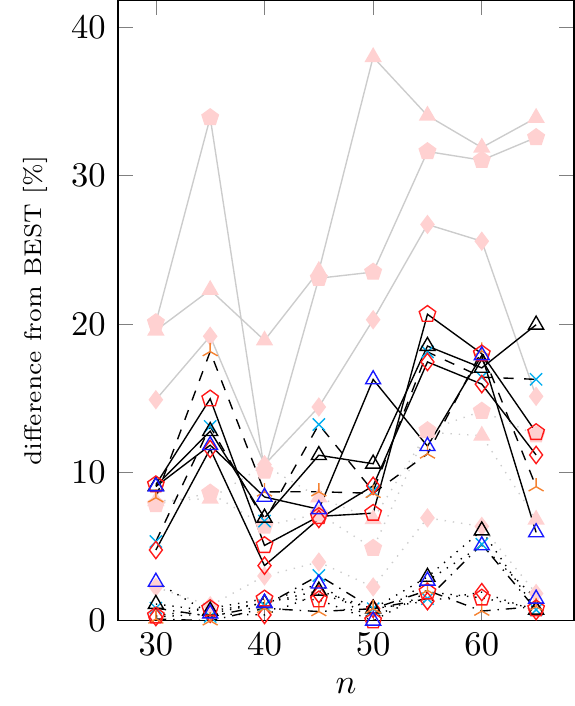}
    \caption{North}
\end{subfigure}%
\caption{Comparison of high-solution-quality heuristics (with a single or 50
    permutations) on the Rome and North instances.}
\label{fig:highperforming_comparison}
\end{figure}

Overall, permutations can significantly improve the results of \emph{mim},
\emph{ccm}, and \emph{plm-none} at the cost of little additional time.
However, when more time is available, \emph{plm} with postprocessing is clearly
preferable.
Multiple permutations of \emph{all} and \emph{inc} can be of use if
one tries to marginally improve already good solutions.
Among the \emph{srm}-algorithms, the relative improvement via permutations is
consistently low with little variance; for a comparison with
the respective \emph{plm}-variants see Figure~\ref{fig:relative_comparison}.
The one outlier is \emph{ccm-srm}, which achieves the greatest relative
improvements for 50 permutations.
Note, however, that we initialize all permutations of \emph{ccm-srm} with a
fixed small chordless cycle instead of a fixed maximal planar subgraph.
This allows for greater variance in the solutions of \emph{ccm-srm} and makes it
difficult to compare the results to other \emph{srm}-algorithms.

The general trend of high-solution-quality algorithms, taking multiple
permutations into account, is shown in
Figure~\ref{fig:highperforming_comparison}:
A single permutation of \emph{mim-srm} or \emph{ccm-srm} will yield better
solutions than a \emph{plm}-variant with incremental postprocessing (but no
\emph{srm}).
Two layers of postprocessing, i.e., \emph{-all-srm} or \emph{-inc-srm}, improve
the results even more.
Solutions resulting from 50 permutations are in a tier of their own, with
\emph{srm}-heuristics achieving higher quality than those without.
Overall, 50 permutations of \emph{mim-srm} or \emph{ccm-srm} provide some of the
best results while taking a lot less time than other algorithms in
their category.
Consider, e.g., the Rome instances in a 50-permutations setting;
\emph{var-inc-srm} can reduce the average solution quality difference to BEST by
only 1.2\% more than \emph{mim-srm}, but its median running time is ten times as
high.

\FloatBarrier
\section{Conclusion}
\label{sec:conclusion}

Our in-depth experimental evaluation not only corroborates the
results of previous papers~\cite{DBLP:journals/jgaa/ChimaniG12,
DBLP:journals/jgaa/ClancyHN19} but also provides new insights into the
performance of star insertion in crossing minimization heuristics.
We presented the novel heuristic \emph{mim}, which proceeds
similarly to the planarization method but inserts most edges by reinserting one
of their endpoints as a star. %
Whenever neither endpoint is a cut vertex of the initial planar subgraph,
the endpoint can be chosen freely, and our experiments indicate
that reinserting \emph{both} endpoints one after another provides the best results.
In general, \emph{mim} performs better than the basic heuristics
from~\cite{DBLP:journals/jgaa/ChimaniG12, DBLP:journals/jgaa/ClancyHN19} that
have a similarly low running time (i.e., \emph{ccm} and \emph{fix-none}).

A central observation is that postprocessing via star insertion
(\emph{srm}) can greatly improve the planarizations resulting from fast
heuristics:
\emph{mim-srm}, \emph{ccm-srm}, and \emph{fix-none-srm} are all faster than the
previously best-performing heuristic \emph{var-inc} and provide better results.
By inserting multiple adjacent edges at once, star (re-)insertion changes the
planarization and its underlying graph decomposition in a way that is sufficient
to properly explore the search space and find good solutions.
Fixed embedding star insertion is thus preferable over the much
slower insertion of edges (or even stars) in a variable embedding setting.

We note that many heuristics---in particular those without edge-wise
post\-processing---are prone to create non-simple crossings (due to lack of
space see
\inappendix{Appendix~\ref{sec:eval_nonsimple_crossings}}%
{\cite[Appendix~C]{chimani2021starstruck}}).
Such crossings can be detected and it is worthwhile to remove them in order to
speed up the procedure and improve the results.
Lastly, multiple permutations are beneficial for heuristics that already
employ postprocessing.
In particular, their application to \emph{mim-srm} and \emph{ccm-srm} provides
very high solution quality at moderate running times.

\bibliography{main}

\inappendix{%
\newpage
\appendix
\section{Statistics on Instance Sets}
\label{sec:instance_statistics}

Figures~\ref{fig:instances_legend}--\ref{fig:complete_instances} give an overview of
the size of the instances mentioned in Section~\ref{sec:experiments}, including
the number of edges that are deleted to create the planar subgraph.
The plots also display the number of these deleted edges that are incident to
one or two cut vertices of the planar subgraph as these are of particular
relevance for the mixed insertion method (see
Subsection~\ref{subsec:heuristics}).
For plots over $n$ or the density of the instance, points for the x-axis values
$x$, $y$ and $z$ represent the mean over all instances with x-axis values in the
intervals $(0,x]$, $(x,y]$ and $(y,z]$.%

\begin{figure}[h]
    \centering
    \includegraphics[scale=1.0]{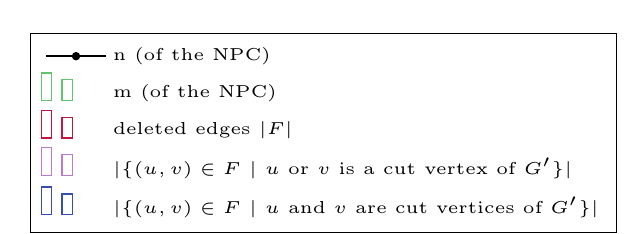}
    \caption{Legend for plots displaying instance statistics.}
    \label{fig:instances_legend}
\end{figure}

\begin{figure}[ht]
\captionsetup[subfigure]{justification=centering}
\centering
\begin{subfigure}[c]{\textwidth}
    \centering
    \includegraphics{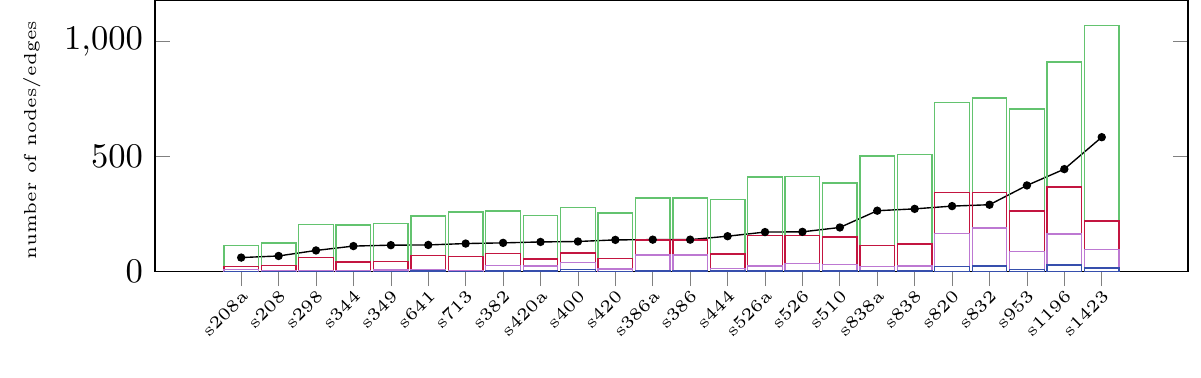}
    \caption{ISCAS-89}
\end{subfigure}\\
\begin{subfigure}[c]{0.45\textwidth}
    \centering
    \vspace*{0.175cm}
    \includegraphics{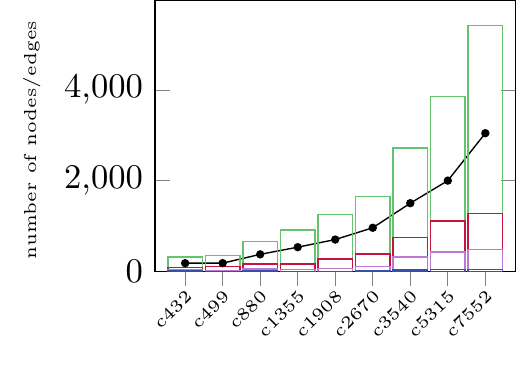}
    \caption{ISCAS-85}
\end{subfigure}\hfill
\begin{subfigure}[c]{0.55\textwidth}
    \centering
    \includegraphics{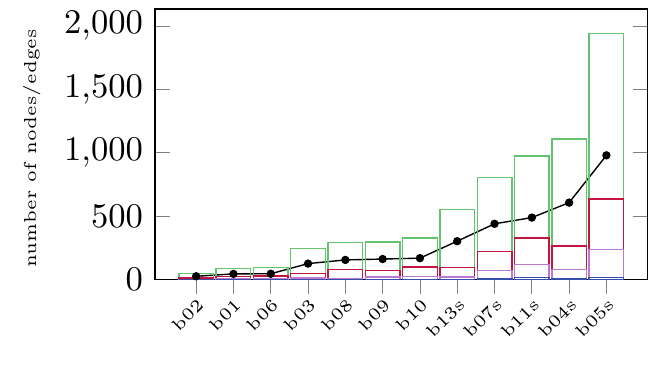}
    \caption{ITC-99}
\end{subfigure}
\caption{Statistics on circuit-based instances.}
\label{fig:circuit_instances}
\end{figure}

\begin{figure}[t]
\captionsetup[subfigure]{justification=centering}
    \centering
        \begin{subfigure}[c]{0.5\textwidth}
            \centering
            \includegraphics{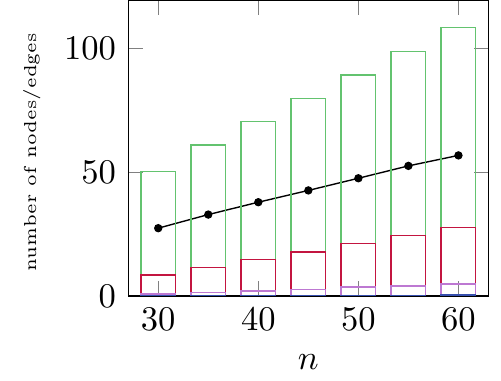}
            \caption{Rome}
        \end{subfigure}\hfill
        \begin{subfigure}[c]{0.5\textwidth}
            \centering
            \includegraphics{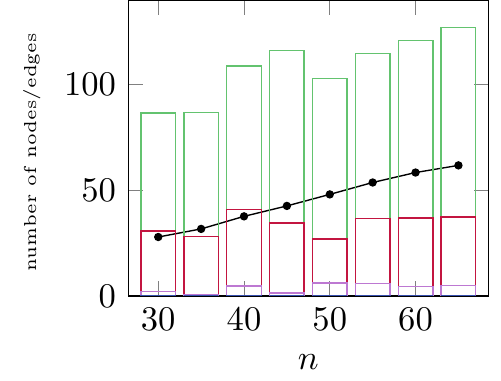}
            \caption{North}
        \end{subfigure}\\[0.3cm]
        \begin{subfigure}[c]{0.4\textwidth}
            \centering
            \includegraphics{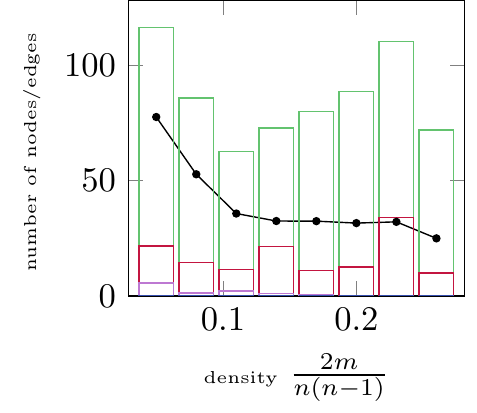}
            \vspace*{0.14cm}
            \caption{Webcompute}
        \end{subfigure}\hfill
        \begin{subfigure}[c]{0.6\textwidth}
            \centering
            \includegraphics{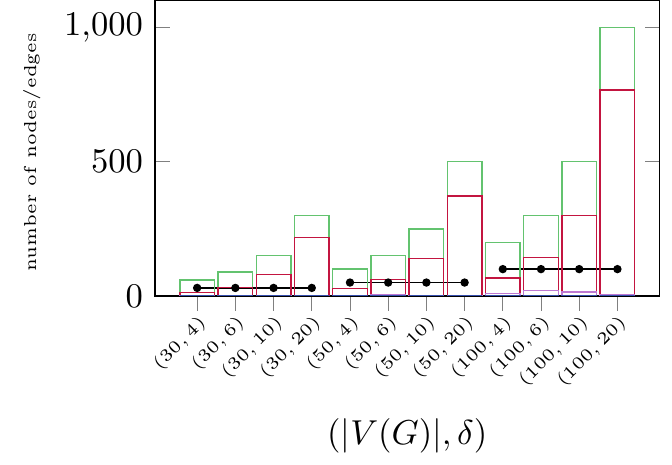}
            \caption{Expanders}
        \end{subfigure}\\[0.3cm]
        \begin{subfigure}[c]{\textwidth}
            \centering
            \includegraphics{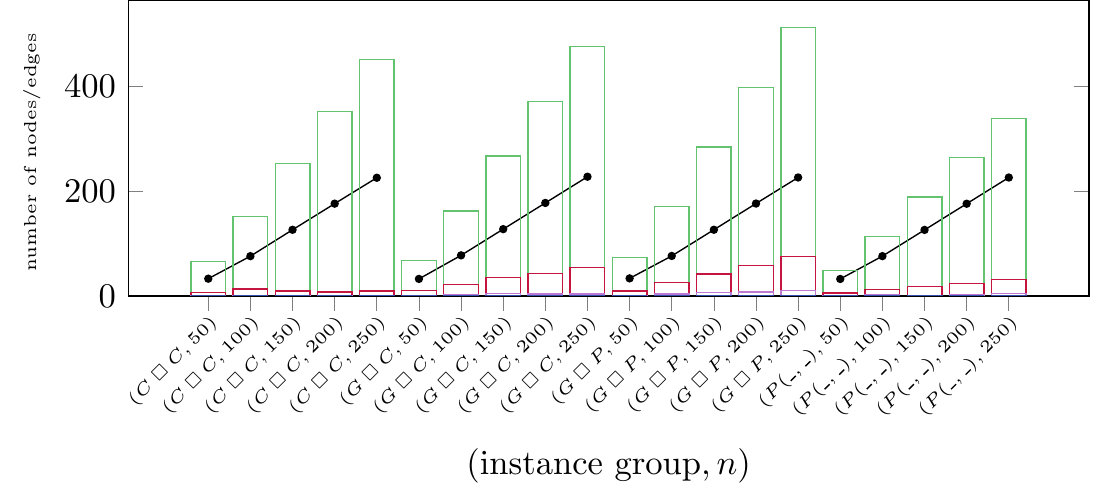}
            \caption{KnownCR}
        \end{subfigure}
        \caption{Statistics on other instances.}
    \label{fig:instance_info}
\end{figure}

\FloatBarrier
\begin{figure}[htp]
\captionsetup[subfigure]{justification=centering}
    \centering
        \begin{subfigure}[c]{0.5\textwidth}
            \centering
            \vspace*{0.25cm}
            \includegraphics{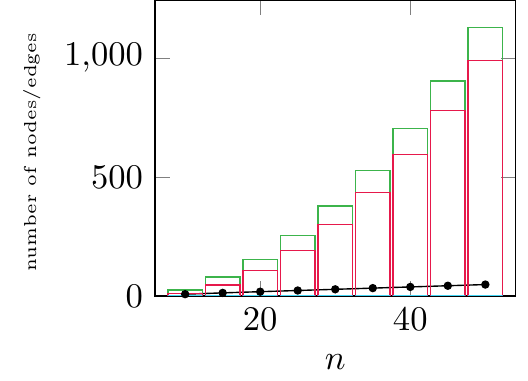}
            \caption{Complete}
        \end{subfigure}\hfill
        \begin{subfigure}[c]{0.5\textwidth}
            \centering
            \includegraphics{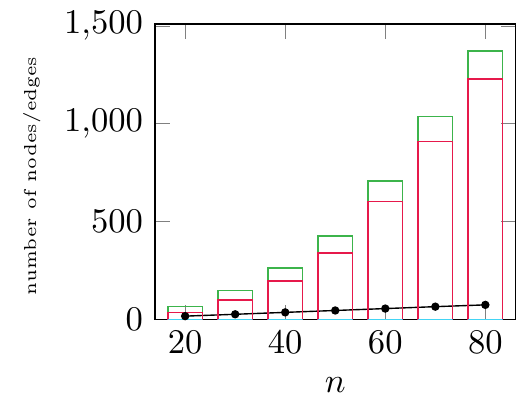}
            \caption{Complete-Bip.}
        \end{subfigure}
        \caption{Statistics on complete (bipartite) instances.}
    \label{fig:complete_instances}
\end{figure}

\section{Planarization Method}
\label{sec:plm_appendix}

This appendix provides an extended version of
Subsection~\ref{subsec:planarization_method}.
Figure~\ref{fig:plm_comparison} showcases the results and running times on the
ISCAS-89 instances, on which \emph{plm} has not been evaluated so far.

\begin{figure}[tb]
\centering
\begin{subfigure}{\textwidth}
    \centering
    \includegraphics{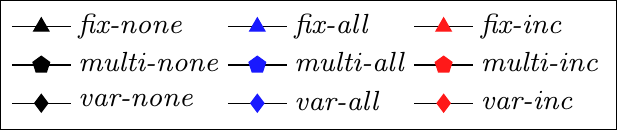}
\end{subfigure}
\\[0.2cm]
\begin{subfigure}{\textwidth}
    \centering
    \includegraphics{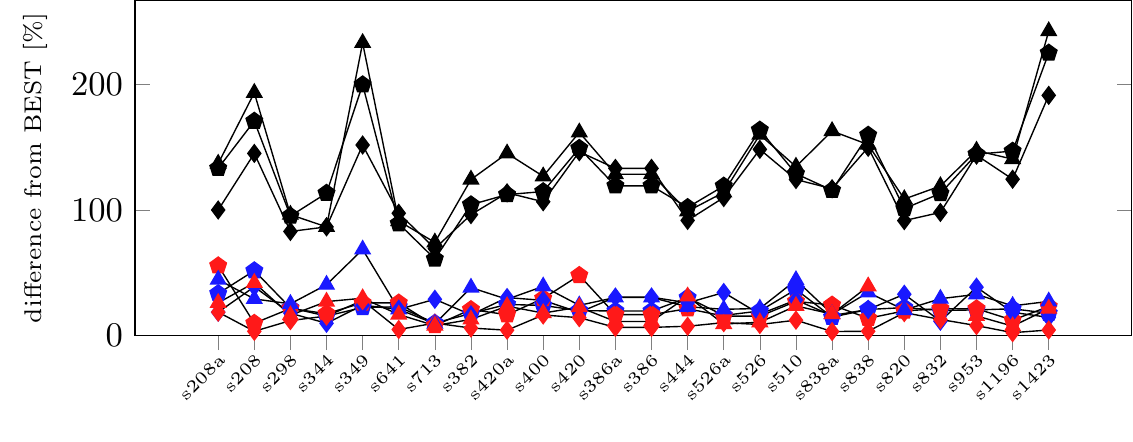}
\end{subfigure}
\\
\begin{subfigure}{\textwidth}
    \centering
    \includegraphics{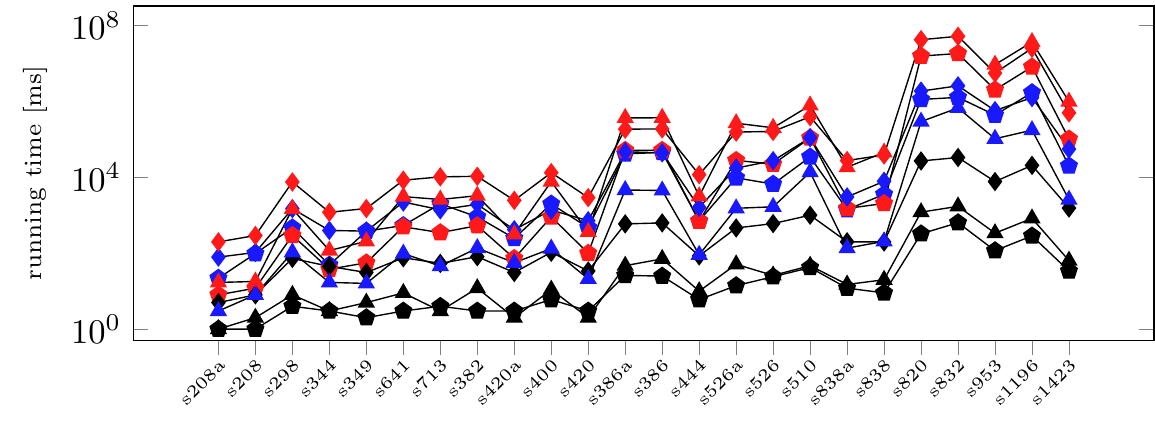}
\end{subfigure}
\caption{Comparison of the \emph{plm}-variants on the ISCAS-89 instances. The
instances s820 and s832 could not be solved with \emph{fix-inc} due to the enforced memory limit.}
\label{fig:plm_comparison}
\end{figure}

One major insight is that (incremental) postprocessing via \emph{all} or \emph{inc}
is a very worthwhile endeavour:
In terms of solution quality, \emph{none} provides much worse results than \emph{all}
and \emph{inc} across all instance sets.
While the switch from \emph{none} to \emph{all} causes a large jump in solution
quality, the improvement of \emph{inc} in comparison to \emph{all} is smaller.
The best results are usually provided by \emph{var-inc}, with \emph{multi-inc}
being a close runner-up.
However, \emph{inc} has the drawback of very high running times and a large
amount of required memory. Several circuit-based instances as well as many
expanders with $|V(G)| \in \{50,100\}, \delta = 20$ could not be solved due to the
enforced memory limit.
For this reason, we did not even attempt to solve the complete (bipartite)
instances with \emph{inc}.

Furthermore, for each of the three postprocessing settings, there is a general
trend among the edge insertion algorithms:
Due to its search across all possible embeddings, \emph{var} performs better
than \emph{multi}, which in turn performs better than \emph{fix}.
This overall hierarchy is especially evident (across all instance sets) when no
postprocessing is employed. On the Rome instances it also persists for
\emph{all} and \emph{inc} respectively.
For other instance sets, the \emph{all}-variants and \emph{fix-inc} produce
planarizations of intermediate quality, but which of these algorithms performs
best depends on the instance.
Here, the prowess of \emph{multi} shines through:
\emph{multi-all} comes close to the solution quality of \emph{var-all} and
sometimes even surpasses it, as is the case on complete instances.
For KnownCR, \emph{multi-all} even trumps~\emph{fix-inc}.

When it comes to running times, \emph{var} is the slowest edge insertion
algorithm as it builds up a new SPR-tree after each edge insertion,
and postprocessing takes a lot of additional time (\emph{inc} especially so).
In some cases, more involved postprocessing can be counter-balanced
by employing fixed embedding edge insertion.
Thus, \emph{fix-all} is faster than \emph{var-none} on the expanders, KnownCR,
and Webcompute instances; and \emph{fix-inc} is faster than \emph{var-all} on Rome and
larger KnownCR instances.
Overall, \emph{fix-all} is the fastest \emph{plm}-variant that still benefits
from the quality improvements of postprocessing.

However, the central observation is the speed of \emph{multi}, which matches the
speed of \emph{fix} in many cases.
On all instance sets except for Rome, \emph{multi-none} becomes even faster than
\emph{fix-none} when considering larger instances.
Furthermore, \emph{multi-inc} is faster than \emph{fix-inc} for Rome,
Webcompute, the circuit-based instances, and dense expanders.
This might be explained by the fact that intermediate planarizations produced by
\emph{multi} contain less crossings and that there is hence less computational
overhead.
As our implementation of \emph{multi} performs incremental postprocessing in a
fixed embedding setting, \emph{multi-inc} also has a speed advantage over
\emph{var-all}
and even \emph{multi-all} (which uses postprocessing in a variable embedding setting) on
a limited number of sparse instances. %
While \emph{multi} does not lead to the best results (this is achieved by
\emph{var-inc}), it constitutes the best compromise between solution quality and
speed.

\section{Removal of Non-simple Crossings}
\label{sec:eval_nonsimple_crossings}

To gain insight into the effectiveness of the removal of non-simple crossings,
we counted how many of them were detected during each algorithm run.
As one would expect, they occur primarily on dense instances, i.e., dense
expanders and complete (bipartite) instances.
When inspecting the final planarizations produced by \emph{plm}, it becomes
evident that in most cases, the postprocessing strategies \emph{all} and
\emph{inc} already remove all non-simple crossings.
In fact, on KnownCR instances, not a single such crossing can be found in the
final solutions.
When no postprocessing is used, however, we found (and removed) such crossings
for 1027--1159 of all instances, with \emph{fix-none} producing more of them than
\emph{multi-none} and \emph{var-none}.
A maximum of 3410 non-simple crossings was created during the run of
\emph{fix-none} on $K_{39,39}$.

For \emph{mim} and \emph{ccm}, we examine the sum of non-simple crossings found
and removed after the reinsertion of each star.
In comparison to \emph{plm} without postprocessing, these numbers are
considerably lower, presumably because inserting multiple edges via star
insertion reduces the likelihood of producing such crossings.
In particular, when inserting a star as described in
Subsection~\ref{subsec:insertion_problems}, it is not possible to introduce a
non-simple crossing between its edges.
For \emph{mim}, there are 517--855 affected instances overall (depending on the
variant), with a maximum of 1916 non-simple crossings for $K_{40,40}$.
While \emph{ccm} results in similarly high numbers (543 instances), the
corresponding distribution of non-simple crossings among the instance sets is
striking:
The algorithm does not produce \emph{any} such crossings on complete (bipartite)
instances but showcases a high occurrence rate on circuit-based instances,
expanders and especially KnownCR.
The numbers hence coincide with the performance of \emph{ccm} on the respective
instances.

With respect to the \emph{srm}-postprocessing, we count only the non-simple
crossings that are removed during the star reinsertion process, with those of
the initial planarization being already removed.
It can be observed that the totals depend heavily on the quality of the initial
planarization---presumably because the creation of a non-simple crossing
requires a specific configuration of already existing crossings (cf.\
Figure~\ref{fig:crossing_existence}).
Among \emph{srm}-algorithms whose initial planarization is constructed using
\emph{all} or \emph{inc}, \emph{fix-all-srm} is the only one with more than 100
instances being affected by non-simple crossings during the
star reinsertion phase (121 instances).
For the remaining \emph{srm}-algorithms, these numbers are considerably higher:
\emph{fix-none-srm} results in the highest amount of affected instances~(1094),
closely followed by \emph{ccm-srm}~(1024), with \emph{mim-srm} producing roughly
half as many~(536).

In summary, the removal of non-simple crossings can significantly improve the
final results of heuristics that do not employ any kind of postprocessing.
It can also slightly improve intermediate planarizations during \emph{srm},
potentially speeding up the procedure.
However, with more involved postprocessing, non-simple crossings are less
likely to occur, and their removal is hence less effective.

\FloatBarrier
\section{Relative Improvement via Permutations}
\label{sec:relative_improvement}

Table~\ref{tab:50_perm_improvement} lists the average relative
improvements of 50 permutations in comparison to the average permutation for
each of the heuristics and instance sets mentioned in
Subsection~\ref{subsec:permutations}.

\begin{table}[htpb]
    \centering
    \caption{Average relative improvement of 50 permutations compared to the
        average permutation in percent.}
    \label{tab:50_perm_improvement}
    \begin{tabular}{l|rrrr}
        Algorithm            & Rome  & North & KnownCR & Webcompute \\
        \hline
        \emph{fix-none}      & 10.60 & 9.57  & 4.91    & 7.60       \\
        \emph{fix-all}       & 17.20 & 11.98 & 7.28    & 12.39      \\
        \emph{fix-inc}       & 18.65 & 12.46 & 9.99    & 15.23      \\
        \emph{multi-none}    & 13.66 & 11.95 & 6.14    & 10.50      \\
        \emph{multi-all}     & 18.10 & 12.83 & 7.87    & 13.23      \\
        \emph{multi-inc}     & 19.08 & 12.98 & 10.00   & 14.28      \\
        \emph{var-none}      & 12.62 & 11.14 & 4.95    & 9.84       \\
        \emph{var-all}       & 18.44 & 12.83 & 8.64    & 14.08      \\
        \emph{var-inc}       & 16.57 & 11.46 & 8.48    & 12.43      \\
        \emph{mim}           & 14.73 & 12.83 & 9.41    & 14.05      \\
        \emph{ccm}           & 34.13 & 27.28 & 27.77   & 33.16      \\
        \emph{fix-none-srm}  & 13.49 & 8.80  & 5.32    & 8.90       \\
        \emph{fix-all-srm}   & 13.23 & 9.25  & 5.02    & 8.13       \\
        \emph{multi-all-srm} & 12.84 & 8.31  & 5.29    & 7.52       \\
        \emph{multi-inc-srm} & 12.93 & 8.93  & 6.72    & 6.90       \\
        \emph{var-all-srm}   & 13.08 & 8.94  & 5.31    & 7.69       \\
        \emph{var-inc-srm}   & 12.12 & 8.45  & 6.50    & 6.45       \\
        \emph{mim-srm}       & 13.59 & 8.77  & 5.63    & 7.81       \\
        \emph{ccm-srm}       & 16.88 & 11.03 & 15.19   & 10.38      \\
    \end{tabular}
\end{table}
}{}

\end{document}